\documentclass{IEEEtran}

\usepackage{cite}
\usepackage{amsthm}
\usepackage{amsmath,amssymb,amsfonts}
\usepackage{bbm}
\usepackage{graphicx}
\usepackage{subfigure}
\usepackage{multirow}
\usepackage{textcomp}
\usepackage[short]{optidef}

\newcommand\Tx{\mathrm{Tx}}
\newcommand\Rx{\mathrm{Rx}}
\newcommand\R{\mathrm{R}}
\newcommand\TR{\mathrm{TR}}
\newcommand\PBF{\mathrm{PBF}}

\allowdisplaybreaks
\begin{document}

\title{An RIS-enabled Time Reversal Scheme\\for Multipath Near-Field Channels}

\author{Andreas Nicolaides, \textit{Member, IEEE,}
	Constantinos Psomas, \textit{Senior Member, IEEE,}\\
	and~Ioannis Krikidis, \textit{Fellow, IEEE}
	\thanks{This work received funding from the European Research Council (ERC) under the European Union's (EU's) Horizon 2020 research and innovation programme (Grant agreement No. 819819) and the EU's Horizon-JU-SNS-2023 research and innovation programme under iSEE-6G (Grant agreement No. 101139291). It was also funded from the EU Recovery and Resilience Facility of the NextGenerationEU instrument, through the Research and Innovation Foundation under RISE (Grand Agreement: DUAL USE/0922/0031).}
	\thanks{The authors are with the IRIDA Research Centre for Communication Technologies, Department of Electrical and Computer Engineering, University of Cyprus, Cyprus (e-mail: \{anicol09, psomas, krikidis\}@ucy.ac.cy).}\vspace{-5mm}
}
\maketitle

\begin{abstract}
	 Time reversal (TR) is a promising technique that exploits multipaths for achieving energy focusing in high-frequency wideband communications. In this letter, we focus on a TR scheme facilitated by a reconfigurable intelligent surface (RIS) which, due to the higher frequency and large array aperture, operates in the near-field region. The proposed scheme enriches the propagation environment for the TR in such weak scattering conditions and does not need channel knowledge for the RIS configuration. Specifically, the RIS is employed to create multiple virtual propagation paths that are required to efficiently apply the TR. We derive a performance bound for the proposed scheme under near-field modeling through the received signal-to-noise ratio (SNR) and we examine how various system design parameters affect the performance. We observe that a linear RIS topology maximizes the number of resolvable paths. It is also demonstrated that the proposed scheme improves the SNR, while for a large number of elements it can outperform the conventional passive beamforming at the RIS.
\end{abstract}

\vspace{-2mm}
\section{Introduction}
	The rapid surge in wireless devices has led to an unprecedented amount of data traffic that the upcoming sixth generation (6G) of wireless networks needs to efficiently handle. In order to address these challenges, a paradigm shift towards higher-frequency communications with wider bandwidths has been proposed \cite{alex2022}. However, as bandwidth increases, more multipaths are resolved, which may cause destructive interference at the receiver \cite{tse2005}. In contrast to most technologies trying to alleviate the negative effects of multipath propagation, time reversal (TR) is an appealing solution that exploits the multipath propagation environment by treating each path as a virtual antenna \cite{alex2022}. In principle, TR is a signal processing technique that uses the time reversed impulse response of the multipath channel as a prefilter at the transmitter. Assuming channel reciprocity, with this technique the energy is focused at the intended receiver in both space and time domains, referred as the \emph{focusing effect} \cite{wand2011}.
	
	Although the TR technique was initially introduced for acoustic communications \cite{fink1992}, its potential performance benefits in wireless communications have attracted significant academic and industrial interest \cite{nguyen2006,wand2011,han2016}. Specifically, in \cite{nguyen2006} the applicability of the TR was experimentally tested for ultra wideband communications, where it was indicated that the TR results in reduced complexity and increased system's capacity. The authors in \cite{wand2011} demonstrated that the TR-based transmission scheme reduces the power consumption significantly and achieves better interference alleviation compared to direct transmission using a conventional Rake receiver. Moreover, in \cite{han2016} it was shown that the TR system, under a sufficiently large bandwidth, can obtain similar performance gains as massive multiple-input multiple-output (MIMO) systems with only a single antenna at the transmitter.
	
	However, it has been shown that the TR-based networks can only provide sufficient performance gains within a rich scattering environment \cite{han2016}. Under weak propagation conditions \textit{e.g.}, due to severe path loss or high blockage sensitivity, the employment of a reconfigurable intelligent surface (RIS) is an appealing solution to overcome this bottleneck. An RIS is a planar metasurface consisting of a large number of passive elements, which are able to individually reflect the incident signals, essentially creating multiple virtual paths in a controllable manner \cite{bjornson2020}. Nevertheless, due to the shift of future 6G networks towards higher frequencies, a large RIS needs to be utilized to obtain sufficient performance gains. As such, the operating regime of the RIS changes from the traditional far-field to the near-field region, which affects the modeling of the wireless channel \cite{bjornson2020,liu2023}.
	
	Motivated by the above, in this paper we propose an RIS-enabled TR scheme, where the RIS, operating in the near-field region, is specifically used to provide additional propagation paths required for improving the focusing accuracy of the TR scheme. To our best knowledge, this is the first work that proposes the RIS employment for facilitating the TR technique. The considered scheme has low implementation complexity, since accurate channel state information (CSI) knowledge is not required for the RIS configuration. By considering near-field channels, we study the received signal-to-noise ratio (SNR) and we provide useful insights on the effect of several design parameters on the system's performance. It is shown that, by increasing the elements more scattering paths can be obtained, while for a large number of elements and a linear RIS topology the proposed scheme achieves higher SNR gains, compared to the passive beamforming (PBF) counterpart.

\vspace{-2mm}
\section{System Model} \label{TR_based_system}

\begin{figure}
		\centering
			{\subfigcapskip=-13pt \subfigure[]{\hspace{9mm}\includegraphics[width=0.82\linewidth]{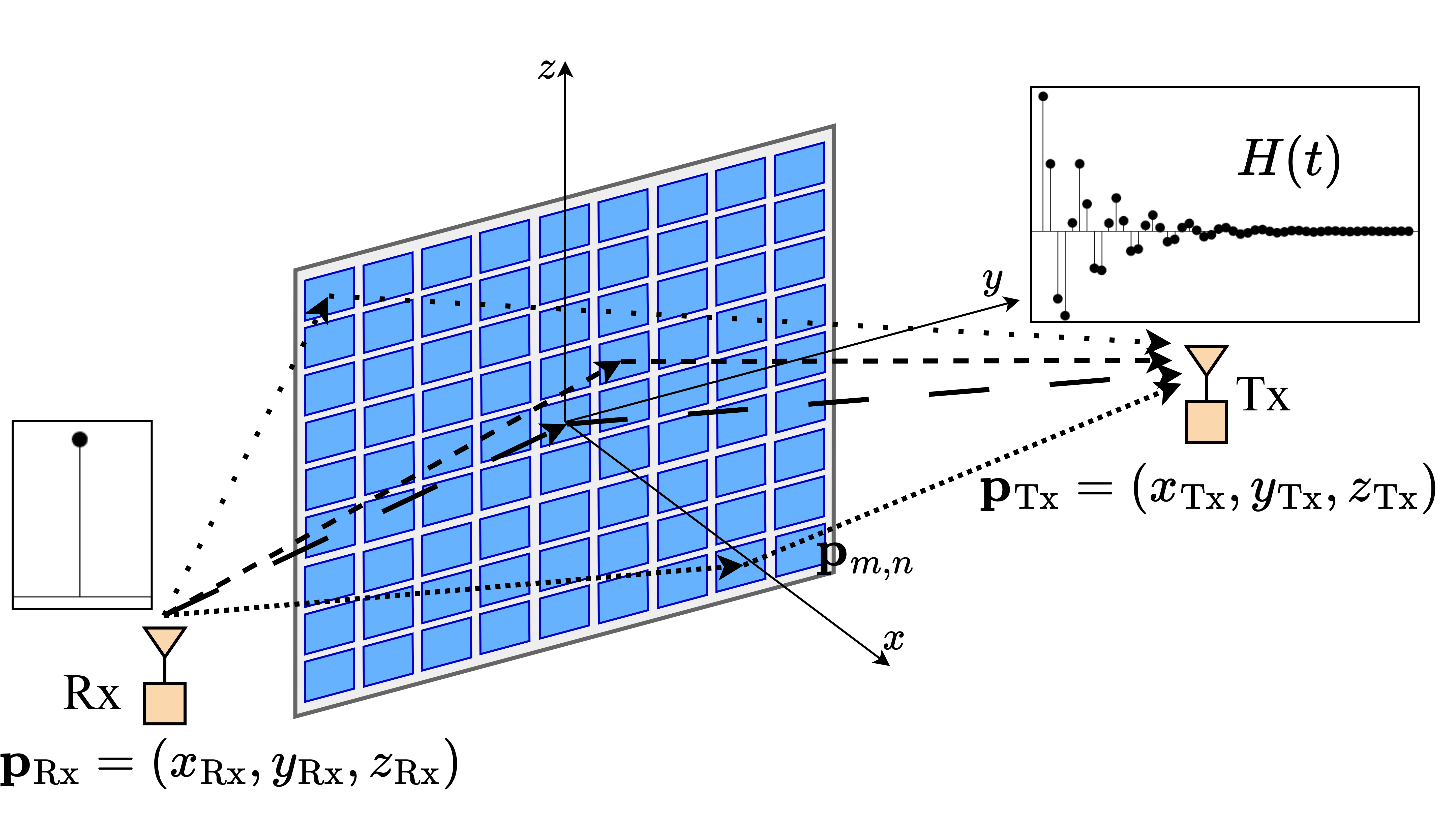}
			\label{fig:syst_a}}}\\\vspace{-5mm}
			{\subfigcapskip=-3pt
			\subfigure[]{\includegraphics[width=0.9\linewidth]{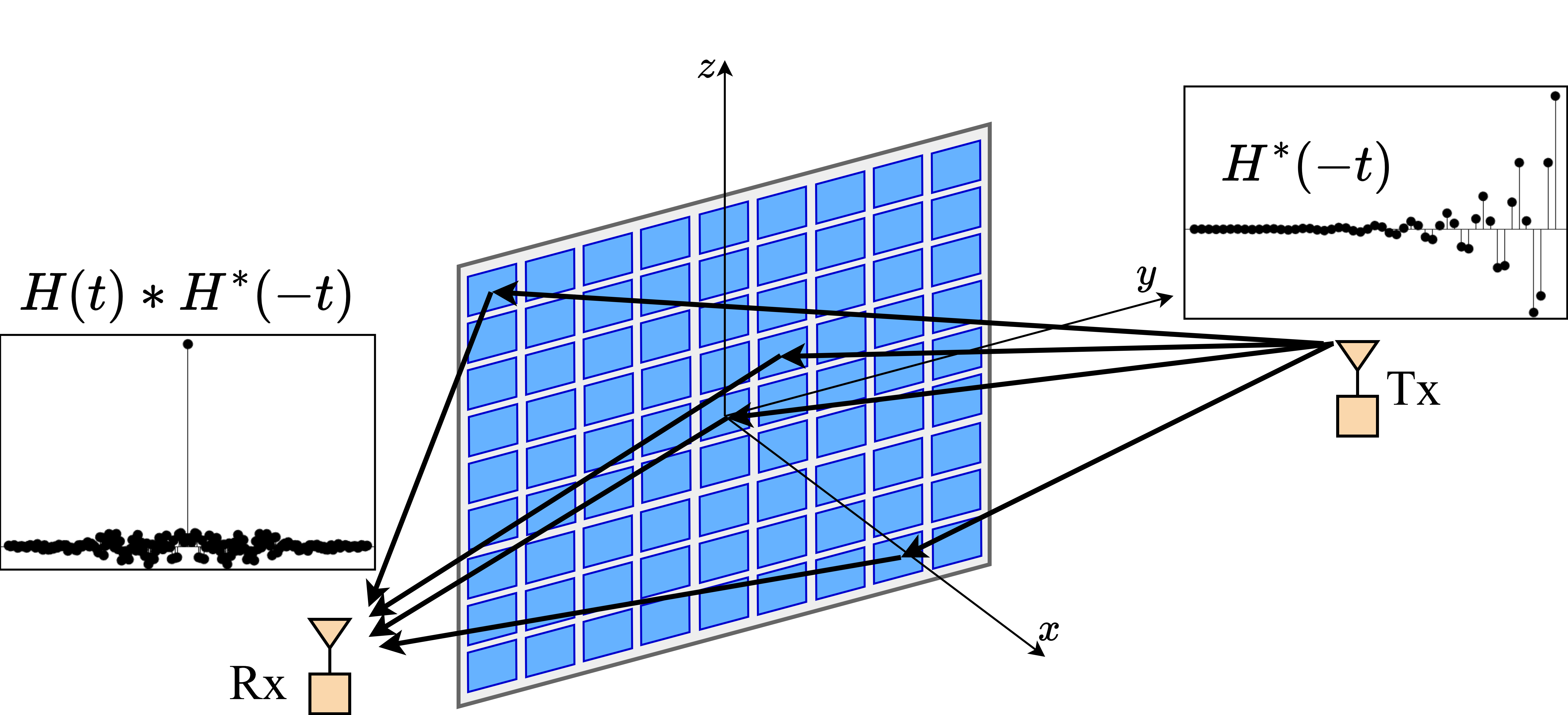}
			\label{fig:syst_b}}}\vspace{-2mm}
		\caption{Network topology and communication procedure of the considered RIS-enabled TR scheme: a) channel probing phase; b) data transmission phase.}\vspace{-5mm}
		\label{fig:system_modelnew}
	\end{figure}

	We consider an RIS-aided communication system shown in Fig. \ref{fig:system_modelnew}, where the communication between a single-antenna transmitter (Tx) and a single-antenna receiver (Rx) is assisted by the employment of an RIS. The communication occurs in the near-field communication regime of the RIS, where the wireless links are considered to be predominantly line-of-sight (LOS) \cite{bjornson2020}. We assume that the direct LOS link between the Tx and Rx is not available (\textit{e.g.}, is blocked by obstacles \cite{bjornson2020}).
	
	The RIS is placed in the $yz$-plane of a Cartesian coordinate system with its geometric center at the origin and consists of $Q$ elements distributed in $M$ rows and $N$ columns \textit{i.e.}, $Q=MN$. For the sake of symmetry, and without loss of generality, we assume the parameters $M$ and $N$ to be odd numbers. In this way, the center element of the RIS corresponds to the origin of the Cartesian system with coordinates $(0,0,0)$. Moreover, the distance between two adjacent elements is considered to be $d=\frac{\lambda}{2}=\frac{c_0}{2f_c}$, where $\lambda$ is the signal wavelength, $f_c$ is the carrier frequency and $c_0$ denotes the speed of light. Therefore, the coordinates of the element located in the $m$-th row and $n$-th column of the RIS are denoted by $\mathbf{p}_{m,n}=(0,nd,md)$, where $m=0,\pm 1, \dots, \pm (M-1)/2$ and $n=0,\pm 1, \dots, \pm (N-1)/2$, respectively. We denote by $\Phi_{m,n}=a_{m,n}\exp\left(\jmath\phi_{m,n}\right)$ the reflection coefficient of the $(m,n)$-th RIS element, where $a_{m,n}\in[0,1]$ is the reflection amplitude, $\phi_{m,n}\in\left[0,2\pi\right)$ is the induced phase shift and $\jmath=\sqrt{-1}$ is the imaginary unit.
	 
	The Tx transmits a passband signal $x_{p}(t)$ derived from the baseband equivalent signal \cite{tse2005}
	\begin{equation}
		x(t)=\sum_{k=0}^{\infty}x[k]g\left( t-\dfrac{k}{W}\right),
	\end{equation}
	where $x[k]\in\mathbb{C}$ denotes the sequence of symbols to be transmitted with constant power $P$, and $g(t)$ is a pulse shaping filter limited by a finite bandwidth $W$. The Tx and Rx are located in the positive direction of the $x$ axis at positions $\mathbf{p}_{\Tx}=(x_{\Tx},y_{\Tx},z_{\Tx})$ and $\mathbf{p}_{\Rx}=(x_{\Rx},y_{\Rx},z_{\Rx})$ respectively, with their corresponding Euclidean distances from the origin given by $\left\| \mathbf{p}_{i}\right\|=\sqrt{x_{i}^2+y_{i}^2+z_{i}^2}, i\in\{\Tx,\Rx\}$. Accordingly, the Euclidean distance between the Tx or the Rx and the $(m,n)$-th element is equal to
	\begin{align}
		r_{i,m,n}&\triangleq\left\|\mathbf{p}_{i}-\mathbf{p}_{m,n} \right\|\nonumber\\
		&=\sqrt{x_i^2+(y_i-nd)^2+(z_i-md)^2}.
	\end{align}
	Note that, since we focus on a near-field communication scenario, the distance of the Tx and Rx from the center of the RIS is upper bounded by the Rayleigh distance $d_{\R}=\frac{2D^2}{\lambda}$ \cite{bjornson2020,liu2023}, where $D$ denotes the largest distance between two elements of the RIS \textit{i.e.}, $D=d\sqrt{(M-1)^2+(N-1)^2}$. Hence, for $i\in\{\Tx,\Rx\}$ we have 
	\begin{equation}
		r_{i,0,0}=\left\| \mathbf{p}_{i}\right\|\leq d\left[ (M-1)^2+(N-1)^2\right].
	\end{equation}
	Finally, we assume that in the considered scenario the near-field LOS channel follows the uniform spherical wave (USW) model \cite{liu2023}. Under this model, the channel coefficients of the system have uniform channel gains based on the free-space path loss. Specifically, the channel coefficient between the Tx or the Rx and the $(m,n)$-th element of the RIS is given by
	\begin{equation}
		h_{i,m,n}=\dfrac{1}{\left\| \mathbf{p}_{i}\right\|\sqrt{4\pi}}\exp\left(-\jmath \frac{2\pi r_{i,m,n}}{\lambda}\right).
	\end{equation}
	In what follows, we demonstrate how the RIS can be utilized as an enabler of the TR scheme, by generating a multipath propagation environment.
	
\vspace{-2mm}
\section{RIS-enabled TR scheme}
	We now proceed to the description of the proposed RIS-enabled TR scheme and evaluate its performance in terms of the signal-to-interference-plus-noise ratio (SINR). As previously mentioned, the main idea of the TR technique is to exploit the scattering environment in such a way that will allow energy focusing in both time and space domains. In order to efficiently apply the TR in the near-field region, we first need to guarantee that a rich scattering environment is available \cite{han2016}. In such weak scattering conditions, the RIS can be employed to enrich the propagation environment since the property of channel reciprocity, which is a fundamental requirement for the TR scheme to work, also holds for the cascaded Tx-RIS-Rx channel \cite{tang2021}. In principle, the TR scheme consists of two phases as shown in Fig. 1: (1) the \textit{channel probing phase}, which is required to provide knowledge regarding the system's multipath environment to the Tx, and (2) the \textit{data transmission phase}, which is dedicated to the transmission of the information signal from the Tx to the Rx, taking into consideration the acquired knowledge \cite{wand2011}. In the following subsections, we present how the TR scheme could be implemented in the context of the considered RIS-aided near-field communication system, by providing a detailed description of each phase.
	
	\subsection{RIS-based Tapped Delay Channel}\label{RIS_TDC}
		According to the TR scheme, during the channel probing phase, the Rx sends an impulse-like pilot signal \cite{wand2011}, which is propagated towards the Tx through the RIS (Fig. \ref{fig:syst_a}). The resulting channel impulse response (CIR) $H(t)$ is then received by the Tx. Note that, this phase is necessary so that the Tx can obtain some knowledge regarding the propagation environment, which is indirectly provided through the received CIR. Assuming that in the considered near-field communication scenario the non-LOS (NLOS) paths provide negligible gains compared to the LOS propagation, the transmitted signal is reflected towards the Rx only by the $Q$ elements of the RIS. These elements can equivalently be seen as scatterers providing a total of $Q$ different propagation paths, each of them generating a delayed and attenuated copy of the transmitted waveform. The multipath CIR ensued from the RIS can therefore be expressed by
		\begin{equation}\label{cont_cir}
			H(t)=\sum_{m,n}h_{m,n}\delta(t-\tau_{m,n}),\vspace{-1mm}
		\end{equation}
		where $h_{m,n}$ and $\tau_{m,n}$ denote the end-to-end channel coefficient and time delay, respectively, of the path occurred by the $(m,n)$-th element. Based on the USW model, the end-to-end channel coefficient of the $(m,n)$-th path is given by\vspace{-1mm}
		\begin{multline}\label{mult_ch}
			h_{m,n}=h_{\Tx,m,n}h_{\Rx,m,n}\Phi_{m,n}=\dfrac{a_{m,n}}{4\pi\left\| \mathbf{p}_{\Tx}\right\|\left\| \mathbf{p}_{\Rx}\right\|}\\
			\times\exp\left[\jmath\left(\phi_{m,n}-\dfrac{2\pi\left(r_{\Tx,m,n}+r_{\Rx,m,n}\right) }{\lambda}\right)\right].
		\end{multline}
		Furthermore, the time delay of the $(m,n)$-th path is equal to\vspace{-1mm}
		\begin{align}
			\tau_{m,n}&=\dfrac{1}{c_0}\left(r_{\Tx,m,n}+r_{\Rx,m,n}+\frac{\lambda\phi_{m,n}}{2\pi} \right)\nonumber\\
			&\approx\dfrac{1}{c_0}\left(r_{\Tx,m,n}+r_{\Rx,m,n} \right).
		\end{align}
		The approximation holds since the delay induced by the phase shift of the RIS element is much smaller than the propagation delay of the corresponding path, hence its impact is negligible.
		
		\begin{figure}[t]
			\hspace{3mm}
			\includegraphics[width=0.75\linewidth]{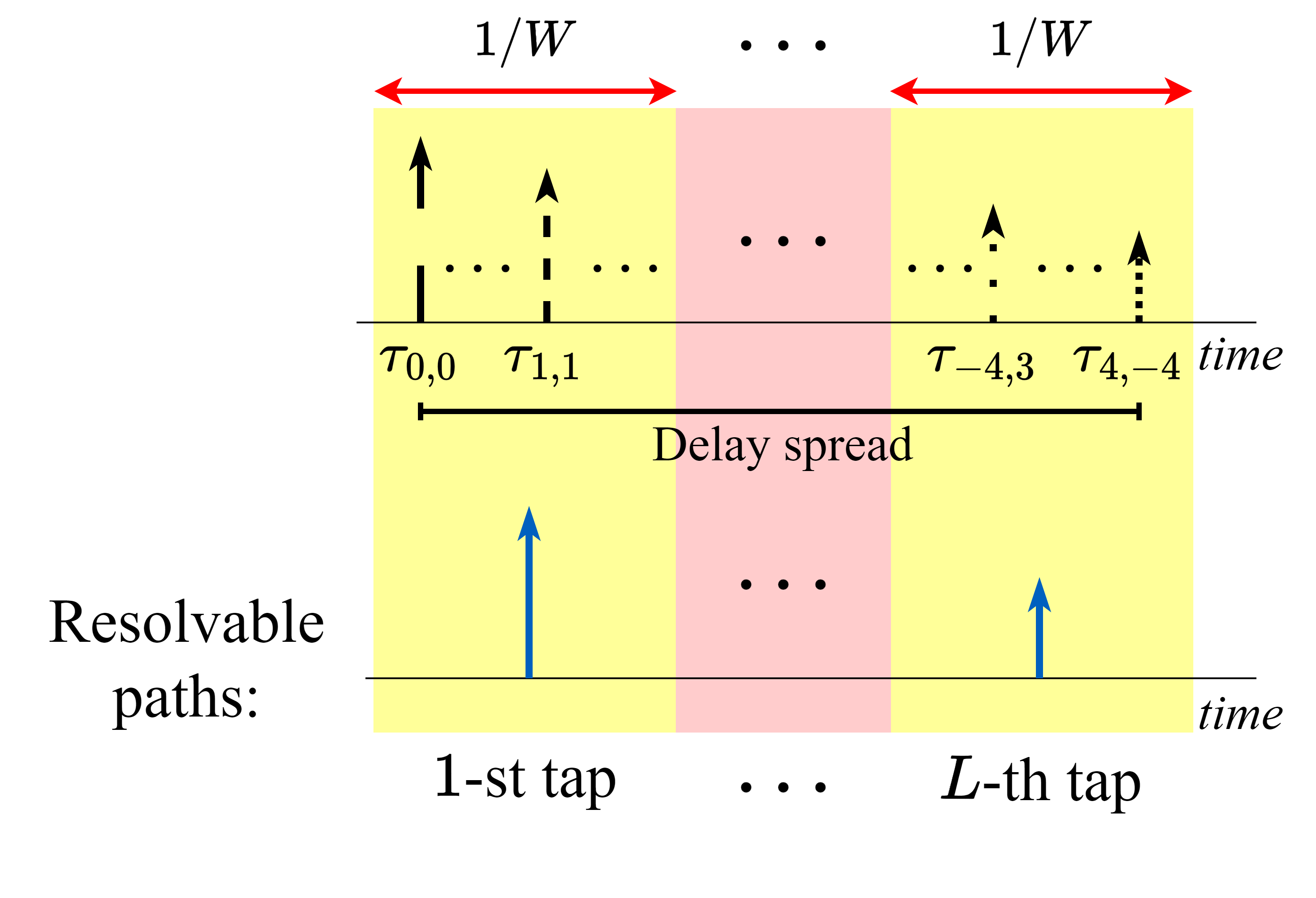}\vspace{-5mm}
			\caption{The RIS-based tapped delay channel.}
			\label{fig:RIStapped}\vspace{-5mm}
		\end{figure}	
		
		However, it is generally hard to retrieve the multipath CIR shown in \eqref{cont_cir}, since the communication system is limited by the available bandwidth $W$, which affects the time delay resolution of the system \cite{tse2005}. Specifically, the resolution to distinguish two propagation paths with different time delays is limited to $1/W$, as shown in Fig. \ref{fig:RIStapped}. The propagation paths whose time delays are within a symbol duration $1/W$ are merged into a single tap. Therefore, the equivalent CIR of the considered system can be modeled as a tapped delay channel with $L$ taps, where each tap corresponds to a group of closely spaced paths that cannot be resolved \cite{tse2005}. Let $\tau_{min}=\min_{m,n} \tau_{m,n}$ and $\tau_{max}=\max_{m,n} \tau_{m,n}$ denote the minimum and maximum delay occurred by the RIS-enabled propagation paths, respectively. Also let $T_{l}$ denote the set of paths that belong to the $l$-th tap, $l=1,\ldots,L$, defined as
		\begin{equation}
			T_{l}=\left\lbrace (m,n): \dfrac{l-1}{W}\leq \tau_{m,n}-\tau_{o} < \dfrac{l}{W}\right\rbrace,
		\end{equation}
		where $\tau_{o}=\lfloor\tau_{min}W\rfloor/W$ indicates the initial time instance of the first tap containing at least one propagation path. The equivalent CIR with limited bandwidth $W$ can therefore be expressed in a simplified discrete time form as
		\begin{equation}
			\label{eqCIR}
			H_{eq}[k]=\sum_{l=1}^{L}h_{eq,l}\delta[k-l], k=1,\ldots,L,\vspace{-1mm}
		\end{equation}
		where the total number of taps is $L=\lceil(\tau_{max}-\tau_{o})W\rceil$,
		and $h_{eq,l}$ contains the channel coefficients of all the paths included in the $l$-th tap, which is equal to\vspace{-1mm}
		\begin{equation}\label{tap_ch}
			h_{eq,l}=\sum_{(m,n)\in T_l}h_{m,n}.\vspace{-1mm}
		\end{equation}
		Note that, if the total delay spread is within a symbol duration, the equivalent CIR is frequency-flat; otherwise, $H_{eq}$ becomes frequency-selective with $L>1$. After receiving the CIR, the system proceeds to the data transmission phase, which will be presented in the next section, along with the analysis for the achieved SINR.\vspace{-3mm}
		
	\subsection{Data Transmission and Achieved SINR}
		In the second phase of the TR scheme, the Tx incorporates the channel knowledge acquired from the channel probing phase into the data sequence in such a way that will enable spatial and temporal focusing of the power at the Rx location. In particular, the Tx time-reverses (and conjugates) the received CIR, and uses the resulting normalized waveform as a prefilter for the information signal (Fig. \ref{fig:syst_b}) \cite{wand2011}. Based on the RIS-enabled equivalent CIR derived in \eqref{eqCIR} for a finite bandwidth $W$, the normalized, time-reversed and conjugate CIR generated by the Tx is given by
		\begin{equation}
			\hat{H}_{eq}[k]=\dfrac{H_{eq}^{*}[L-k]}{\sqrt{\sum_{l=1}^{L}\left|H_{eq}[l] \right|^{2} }}, k=1,\ldots,L.
		\end{equation}
		By embedding the above waveform into the sequence of data symbols $\{x[k]\}$, the transmitted signal can be expressed as the convolution of the time-reversed CIR with the information signal \textit{i.e.}, $s[k]=\sqrt{P}\left( x\ast \hat{H}_{eq}\right) [k]$. Due to the channel reciprocity, the RIS-based multipath channel acts as a natural matched filter to $\hat{H}_{eq}[k], k=1,\ldots,L$. As a result, the convolution of the time-reversed CIR with the multipath channel provides a unique peak at the receiver's location. Note that, the maximum-power peak of the autocorrelation function of the CIR occurs when $k=L$.
		
		We can now proceed to the performance analysis of the TR scheme for the considered system in terms of SINR. The signal received at the Rx at the end of the data transmission phase can be written as the sum of the useful signal and an inter-symbol interference (ISI) term as follows\vspace{-1mm}
		\begin{align}
			y[k]&=\sqrt{P}\left(x\ast \hat{H}_{eq}\ast H_{eq}\right) [k]+n[k]\nonumber\\
			&=\underbrace{\sqrt{P}\left(\hat{H}_{eq}\ast H_{eq}\right)[L]x[k-L]}_{\textrm{Useful signal}}\nonumber\\
			&\quad \underbrace{+\sqrt{P}\sum_{l=1,l \neq L}^{2L-1}\left(\hat{H}_{eq}\ast H_{eq}\right)[l]x[k-l]}_{\textrm{ISI}}+n[k],
		\end{align}
		where $n[k]\sim \mathcal{CN}(0,\sigma^2)$ is the additive white Gaussian noise with variance $\sigma^2$. The SINR can be expressed as
		
		\begin{equation}
			\gamma_{\TR}=\dfrac{P_\textrm{U}}{P_{\textrm{ISI}}+\sigma^{2}},
		\end{equation}
		where $P_\textrm{U}$ is the power of the useful signal which, by combining \eqref{mult_ch} and \eqref{tap_ch}, is equal to\vspace{-1mm}
		\begin{align}\label{p_use}
			P_\textrm{U}&=P\left|\left(\hat{H}_{eq}\ast H_{eq}\right)[L]\right|^{2}=P\sum_{l=1}^{L}\left|h_{eq,l} \right| ^{2}\nonumber\\
			&=\dfrac{P}{16\pi^{2}\left\| \mathbf{p}_{\Tx}\right\|^{2}\left\| \mathbf{p}_{\Rx}\right\|^{2}}\sum_{l=1}^{L}\bigg|\sum_{(m,n)\in T_l}a_{m,n}\nonumber\\
			&\:\:\:\:\times\exp\left[\jmath\left(\phi_{m,n}-\dfrac{2\pi\left(r_{\Tx,m,n}+r_{\Rx,m,n}\right) }{\lambda}\right)\right] \bigg| ^{2} ,
		\end{align}
		and $P_\textrm{ISI}$ is the ISI power given by\vspace{-1mm}
		\begin{equation}
			P_\textrm{ISI}=P\sum_{l=1,l \neq L}^{2L-1}\left| \left(\hat{H}_{eq}\ast H_{eq}\right)[l]\right| ^{2}.\vspace{-1mm}
		\end{equation}
		Since we study the potential of employing the RIS for facilitating the TR technique, we consider an ideal scenario where the ISI can be fully eliminated; ISI suppression can be achieved through sophisticated signal processing techniques, including up-sampling by a large rate back-off factor \cite{wand2011} or advanced waveform design \cite{han2016}. Thus, we focus only on the achieved SNR $\tilde{\gamma}_{\TR}$, which serves as a useful upper bound on the performance of the RIS-enabled TR scheme \textit{i.e.},\vspace{-1mm}
		\begin{equation}
			\gamma_{{\TR}}\leq \tilde{\gamma}_{\TR}=\dfrac{P_\textrm{U}}{\sigma^{2}}.\vspace{-1mm}
		\end{equation}
		Based on the above expressions, we observe that the performance of the considered RIS-enabled TR system depends on the number of observable taps $L$, as well as the cardinality of the sets $T_{l},1\leq l\leq L,$ \textit{i.e.}, the number of elements within each tap. In what follows, we provide a discussion on how various aspects of the system deployment can affect these parameters, and therefore the SNR achieved by the proposed scheme.
		
	\subsection{Insights into System Design Effects}\label{discussion}
		Since the presented RIS-based TR scheme is highly sensitive to several system design parameters, it is of particular interest to examine how the RIS should be deployed in order to maximize the potential gains provided by the application of this technique in wireless systems. Based on how the RIS-enabled tapped delay channel is generated in Section \ref{RIS_TDC}, the number of taps and the number of non-resolvable paths considered in each tap can be controlled either by modifying the RIS configuration, that is, the total number of elements $Q$ and their deployment into $M$ rows and $N$ columns, or by changing the value of the available bandwidth $W$. In particular, as the number of elements at the RIS increases, the number of observable taps can be increased as well, resulting in an improvement of the obtained SNR. On the other hand, for a fixed number of elements, the RIS topology and the available bandwidth could still affect the system's channel resolution.
		
		For the sake of presentation, Table \ref{Table1} shows the number of resolvable taps that are recorded in an indicative RIS-aided system operating at $f_c=10$ GHz with $Q=1225$ elements over various bandwidth values and for different RIS topologies. We observe that, 
		\begin{itemize}
			\item As the RIS structure changes from the square-shaped RIS towards a stripe-like RIS with $M\ll N$, the resulting channel can be resolved into more taps. In particular, we observe from Table \ref{Table1} that by deploying the RIS as a linear array of elements ($M=1$), $L$ becomes significantly larger compared to other RIS configurations.
			\item For a given RIS topology, increasing the available bandwidth $W$ can also lead to a larger number of taps. However, it is important to note that a larger number of taps does not always improve the achieved SNR. Specifically, although a higher value of $W$ will provide more taps, for a given RIS configuration the number of non-resolvable paths in each tap will decrease, and so the channel gains of the taps will be reduced.
		\end{itemize} 
		\begin{table}[t!]
			\caption{Number of resolvable taps $L$\\ $\mathbf{p}_{\Tx}=(2,2,0),\mathbf{p}_{\Rx}=(2,-2,0), Q=1225, f_{c}=10$ GHz}\label{Table1}
			\centering
			\begin{tabular}{| c | c || c | c |}\hline
				\multicolumn{2}{|c||}{$W=2$ GHz} & \multicolumn{2}{|c|}{$W=4$ GHz}\\\hline
				$M\times N$ & Number of taps ($L$) & $M\times N$ & Number of taps ($L$)\\\hline
				$35\times 35$ & $1$ & $35\times 35$ & $1$\\
				$7\times 175$ & $3$ & $7\times 175$ & $5$\\
				$5\times 245$ & $6$ & $5\times 245$ & $10$\\
				$1\times 1225$ & $89$ & $1\times 1225$ & $176$\\\hline
			\end{tabular}\vspace{-5mm}
		\end{table}
		\noindent It is therefore deduced that, although the proposed scheme can be applied for any system configuration, the deployment of a linear RIS (or stripe-like RIS \cite{ganesan2020}) is beneficial for the efficient application of the TR, since this topology can significantly enrich the scattering environment, by maximizing the number of generated propagation paths. The adoption of this topology has several advantages, such as easier deployment in indoor scenarios \textit{e.g.}, along the perimeter of a large conference room or a shopping mall, while it is also more robust to the presence of obstacles compared to a square-shaped RIS \cite{ganesan2020,dardari2022}.
		
\section{Numerical Results} \label{results}
	Based on the discussion provided in Section \ref{discussion}, in the following figures we focus on the performance of the proposed scheme under a linear RIS topology deployed horizontally ($M=1, N=Q$), in order to maximize the number of obtained paths. For the sake of presentation, we consider a system operating at $f_c=10$ GHz (and thus $d=1.5$ cm), with the positions of the Tx/Rx set to $\mathbf{p}_{\Tx}=(2,2,0)$ and $\mathbf{p}_{\Rx}=(2,-2,0)$, respectively. Moreover, we set $P=30$ dBm and $\sigma^2=1$. Since our focus in the proposed scheme is to employ the RIS for increasing the number of paths and the reflection coefficients of the RIS elements have negligible impact on the derived time delays, we assume $a_{m,n}=1$ and $\phi_{m,n}=0,\:\forall m,n$. Note that, with a different RIS configuration \textit{e.g.}, through co-phasing, the proposed scheme can achieve even higher gains. However, such design requires knowledge of each channel coefficient prior to the channel probing phase.
	
	\begin{figure}[t!]\centering
		\includegraphics[width=0.8\linewidth]{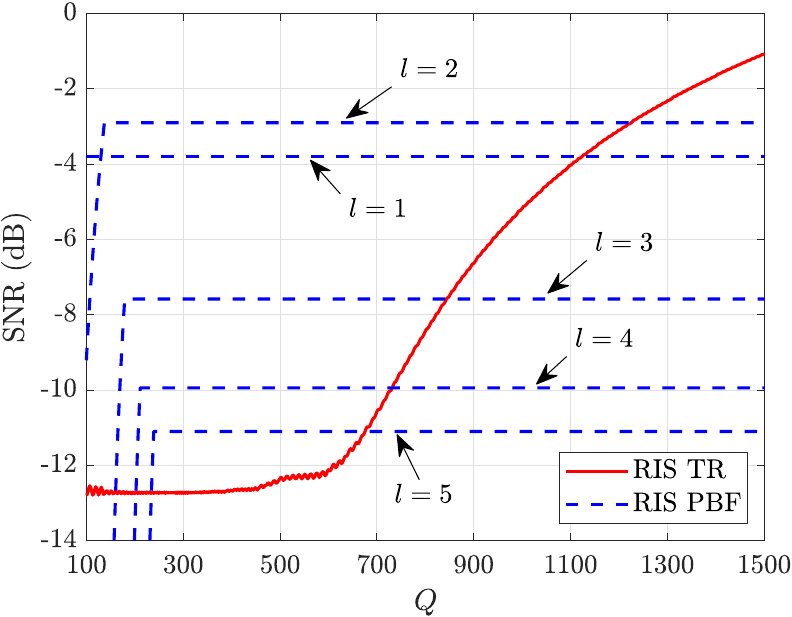}\vspace{-2mm}
		\caption{Achieved SNR under RIS-based TR vs RIS PBF for $W=2$ GHz.}\vspace{-5mm}
		\label{fig:TR_beam}
	\end{figure}
	
	Fig. \ref{fig:TR_beam} shows the achieved SNR of the RIS-enabled TR scheme with respect to the number of elements. The performance of the proposed scheme is also compared to the case where the RIS performs (CSI-based) PBF. Similar to the proposed scheme, we assume ideally that the ISI can be fully eliminated \cite{arslan2021}. By setting the phase shifts of the elements corresponding to the $l$-th tap as $\phi_{m,n}=2\pi\left(r_{\Tx,m,n}+r_{\Rx,m,n}\right)/\lambda$, the SNR for the $l$-th tap is
	\begin{equation}
		\tilde{\gamma}_{l,\PBF}=\dfrac{P\left| T_l\right| ^2}{16\pi^{2}\left\| \mathbf{p}_{\Tx}\right\|^{2}\left\| \mathbf{p}_{\Rx}\right\|^{2}\sigma^{2}},
	\end{equation}
	where $\left| T_l\right|$ denotes the cardinality of the set of paths corresponding to the $l$-th tap. Based on how the RIS-based tapped delay channel is derived, it is observed that the cardinality of each tap, as well as the strongest tap, depends on the number of elements, the bandwidth $W$ and the positions of the Tx and Rx. It can be seen that as the number of elements increases, the performance of the proposed scheme is improved, while for a large number of elements it outperforms the highest SNR gain achieved with RIS PBF. This is due to the fact that by increasing the elements of the RIS, the number of generated taps is also increased, providing a rich scattering environment for the efficient implementation of the TR scheme. On the other hand, under the RIS PBF the SNR for the $l$-th tap depends only on the propagation paths within the specific tap, so further increase of the elements will not improve the performance. Note that, the slight ripple effect observed at the achieved SNR of the proposed scheme is due to the non-monotonic nature of \eqref{p_use}, which as expected disappears in the PBF case.
	
\begin{figure}[t!]\centering
		\includegraphics[width=0.8\linewidth]{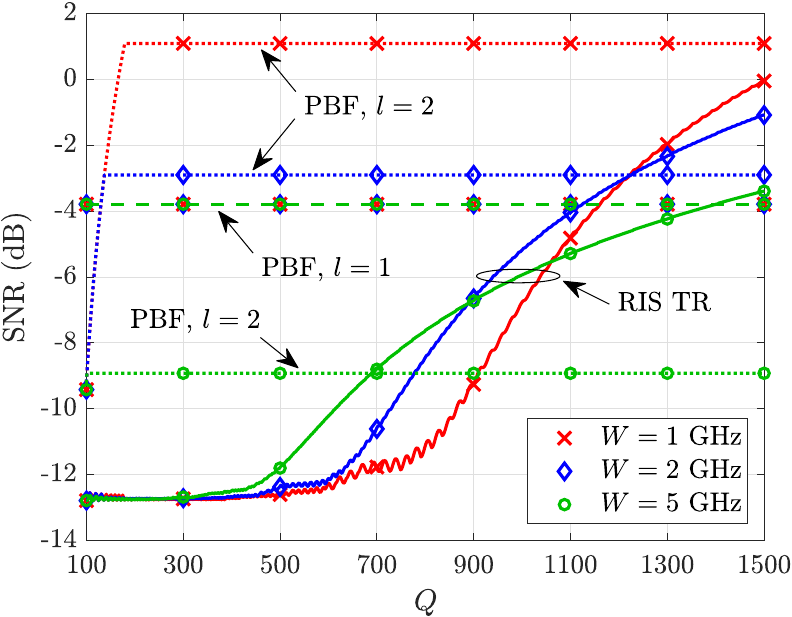}\vspace{-2mm}
		\caption{Achieved SNR versus $Q$ for different values of bandwidth $W$.}\vspace{-5mm}
		\label{fig:TRW}
	\end{figure}

	In Fig. \ref{fig:TRW}, we show how the available bandwidth $W$ affects the performance of the RIS-enabled TR and the RIS PBF with respect to the number of elements. It is observed that, under the RIS-enabled TR, for a moderate number of elements, a larger bandwidth provides a higher SNR gain. As previously mentioned, by considering a larger bandwidth, we can obtain a larger number of taps. Thus, with a smaller number of elements the scattering environment is sufficiently enriched, and the gains of the TR technique can be obtained. However, for a large number of elements, by increasing the bandwidth the performance is deteriorated. Note that, in this case the number of generated taps is sufficiently large under all the considered values of $W$. On the other hand, for a larger bandwidth, each tap contains a smaller number of non-resolvable paths, and so the channel gain obtained for each tap is decreased. Therefore, for a very large number of elements increasing the bandwidth reduces the achieved SNR. Finally, we observe that the SNR obtained under the RIS PBF can be significantly degraded when the available bandwidth is increased. In this case, the proposed scheme can outperform the PBF scenario with a smaller number of elements.
	
\section{Conclusion}
	In this letter, we investigated the potential gains of a TR scheme facilitated by an RIS, which operates in the near-field region. By generating multiple virtual propagation paths, the RIS enriches the scattering environment required for the TR. The performance limits of the proposed scheme were studied through the received SNR and useful guidelines for the design of RIS-based TR systems were provided. Future directions include multi-user scenarios, exploring the effect of ISI on the achieved performance, or the consideration of a metasurface that can control the time delays of the propagation paths.

\end{document}